\documentclass[aps,prd,onecolumn,showpacs,nofootinbib,amsmath,amssymb,floatfix,superscriptaddress,showkeys]{revtex4}
\pdfoutput=1
\usepackage{bm}
\usepackage{graphicx}
\usepackage{amssymb,amsmath}
\usepackage{physics}
\usepackage{multirow}
\usepackage{units,changes}
\usepackage{color,url}
\usepackage{tabu}
\usepackage{placeins}
\usepackage{array}
\usepackage[colorlinks=true,urlcolor=blue,anchorcolor=blue
,citecolor=blue,filecolor=blue,linkcolor=blue,menucolor=blue
,linktocpage=true,pdfproducer=medialab,pdfa=true]{hyperref}
\usepackage{colordvi}
\def\beqa{\begin{eqnarray}}
\def\eeqa{\end{eqnarray}}

\usepackage{soul}
\usepackage{hhline} 

\newcommand{\kev}{\ensuremath{\,\mathrm{keV}}}
\newcommand{\mev}{\ensuremath{\,\mathrm{MeV}}}
\newcommand{\gev}{\ensuremath{\,\mathrm{GeV}}}

\newcommand{\kpc}{\ensuremath{\,\mathrm{kpc}}}

\begin{document}

\title{Probing Cosmic-Ray-Boosted and Supernova-Sourced Sub-GeV Dark Matter with Paleo-Detectors}
\def\slash#1{#1\!\!\!/}

\author{Xiaoyong Chu}
\email{chuxiaoyong@ucas.ac.cn}
\affiliation{International Centre for Theoretical Physics Asia-Pacific (ICTP-AP),
University of Chinese Academy of Sciences, 100190 Beijing, China}

\author{Yue-Lin Sming Tsai}
\email{smingtsai@pmo.ac.cn}
\affiliation{Key Laboratory of DM and Space Astronomy, Purple Mountain Observatory, Chinese Academy of Sciences, Nanjing 210023, China}
\affiliation{School of Astronomy and Space Science, University of Science and Technology of China, Hefei 230026, China}

\author{Meiwen Yang}
\email{mwyang@pmo.ac.cn}
\affiliation{Key Laboratory of DM and Space Astronomy, Purple Mountain Observatory, Chinese Academy of Sciences, Nanjing 210023, China}
\affiliation{School of Astronomy and Space Science, University of Science and Technology of China, Hefei 230026, China}

\begin{abstract}
Astrophysical dark matter particles with masses well below GeV-scale can be difficult to detect using conventional nuclear recoil experiments due to their low velocities in our Milky Way halo. 
Elastic scattering with high-energy cosmic rays  or thermal production inside core-collapse supernovae  can accelerate sub-GeV DM to (semi-)relativistic velocities, 
producing nuclear recoil energies above the keV threshold that paleo-detectors can record over geological timescales.
Using olivine as the target with 100~g$\cdot$Gyr exposure, we compute track length distributions from such (semi-)relativistic dark matter fluxes, incorporating all major backgrounds (neutrinos, uranium-chain neutrons, thorium recoils) with 
a statistical analysis on an Asimov dataset. 
We derive 95\% C.L. projected sensitivity of  paleo-detectors to the DM-nucleon cross section for dark matter masses between a few MeV and hundreds of MeV. 
Our results show that paleo-detectors are able to probe large parameter regions that are not covered by current and near-future  experiments designed to detect dark matter and neutrinos.  
In particular,  paleo-detectors offer a unique ability to record the dark matter flux from Galactic supernova events over geological times. Such cumulative exposure enables sensitivity gains of a few orders of magnitude compared to conventional experiments.
\end{abstract}

\maketitle

\section{Introduction}


Dark matter (DM) constitutes about a quarter of the Universe, yet its particle nature remains unknown. Weakly Interacting Massive Particles (WIMP), together with the so-called WIMP miracle~\cite{Lee:1977ua,Jungman:1995df},  have long been compelling candidates, but experiments have so far only reported null signals~\cite{Chang:2008aa,HESS:2009chc,PAMELA:2008gwm,Hooper:2010mq,Zhou:2014lva,Calore:2014xka,Daylan:2014rsa,Cui:2016ppb,Cuoco:2016eej,Cholis:2019ejx,Fan:2022dck,Fan:2024rcr,Feng:2008ya,Arcadi:2017kky,Cirelli:2024ssz}. 
This has motivated the study of sub-GeV DM, which arises naturally in the thermal freeze-out framework and appears in various new physics models; see e.g. ~\cite{Pagels:1981ke,Dodelson:1993je, Boehm:2002yz,Boehm:2003hm} and more recently \cite{Bertone:2018krk, Matsumoto:2018acr, Binder:2022pmf, Chen:2024njd, An:2024nsw, Watanabe:2025pvc,Tang:2025vqf}. 
However, at the typical Galactic velocity of $10^{-3}c$, sub-GeV DM elastically scattering off heavy target nuclei  produces recoil energies  below the keV threshold of conventional direct detection experiments~\cite{LZ:2018qzl, XENON:2018voc}. 
To address this difficulty, several mechanisms of astrophysically accelerated DM have been proposed, 
including cosmic-ray-boosted DM (CRDM)~\cite{Cappiello:2018hsu,Bringmann:2018cvk, Ema:2018bih, Dent:2019krz,Wang:2019jtk, Guo:2020drq, Guo:2020oum,  Jho:2020sku,Dent:2020syp,Ema:2020ulo, Bell:2021xff,Xia:2021vbz, Feng:2021hyz, Wang:2021nbf,Lu:2023aar, Ghosh:2024dqw, Jeesun:2026lro} and 
supernova-sourced DM (SNDM)~\cite{DeRocco:2019jti, Baracchini:2020owr,Bhalla:2025vnq, Alonso-Gonzalez:2026syw}. 
In the CRDM mechanism, astrophysical DM particles get boosted via elastic scattering with high-energy cosmic ray (CR),  while in the SNDM mechanism, Galactic supernova (SN) events with internal temperatures around $\mathcal{O}(10-100)\ \mev$ thermally produce DM particles with mass below/around  the SN temperatures, generating an (semi-)relativistic DM flux. Through either mechanism, the nuclear recoil signals resulting from energetic dark matter scattering are raised to detectable levels. Nevertheless, the flux of such boosted DM is typically much smaller than that of the virialized DM component inside the Galactic halo, requiring a large exposure mass to achieve meaningful experimental sensitivities. 


Interestingly, compared with direct detection experiments that increase the target mass, using ancient minerals offers a complementary approach. This approach, hereafter called paleo-detectors, has been studied extensively for DM candidates with masses above GeV~\cite{Baum:2018tfw, Edwards:2018hcf, Drukier:2018pdy, Baum:2021chx, Baum:2021jak, Baum:2024eyr}.
Rather than relying on large target masses, they achieve high sensitivities through ultra-long exposure times, retaining damage tracks from DM–nucleus interactions in crystal lattices over billion-year timescales. 
For instance, 100\,g of material recording nuclear recoils for 1\,Gyr provides an effective exposure of $\mathcal{O}(10^5)$~ton$\cdot$yr, 
far exceeding that of direct detection experiments. 
In paleo-detectors, DM particles scatter off target nuclei, producing primary recoil atoms that migrate through the lattice and create damage tracks, which can be read out with X‑ray,  ion‑beam, or electron  microscopy; see \emph{e.g.} \cite{Lider_2017, HILL201265, Calabrese-Day:2026soq}. 
As pointed out in Ref.~\cite{Collar:1995aw}, paleo-detectors suffer from contamination such as 
CRs, underground radioactive decays, and neutrinos, thus the DM signal may lie well below the total backgrounds. 
To reduce these backgrounds, two criteria guide the mineral selection.  First, to mitigate radioactive backgrounds, particularly from $^{238}\mathrm{U}$ decays, ultra-basic rocks, such as olivine $\rm (Fe, Mg)_2SiO_4$, are suitable. This is because ultra-basic rocks originate from the mantle and have uranium concentrations far lower than those of typical crustal minerals~\cite{Drukier:2018pdy}. 
Secondly, to reduce the CR background, the mineral must be obtained from sufficient depths. Concretely, deep borehole cores~\cite{doi:10.1126/science.aar2687, KREMENETSKY198611}  with rock overburden exceeding about 5\,km are effectively shielded from CRs.
In addition to the selection, the track length distribution also helps distinguish DM signals from backgrounds, as demonstrated in Ref.~\cite{Fung:2025cub}.

In this work, we study the projected sensitivities of paleo-detectors in  detecting the scattering cross section between nucleons and boosted sub-GeV DM. 
While this idea has been explored for the  CRDM mechanism mostly within a dark photon mediator model~\cite{Wang:2026you},  
here we use a constant scattering cross section and adopt a much more conservative treatment of the backgrounds by only considering the short-track regime. 
That is, the sensitivities of paleo-detectors to the CRDM mechanism derived in this work are conservative and thus robust. 
Then we additionally consider the SNDM mechanism, in which a (semi-)relativistic DM flux is pair produced by nearby SN events. In this case, the dark flux approximately follows a Fermi-Dirac momentum distribution, peaking around the SN temperature, in contrast to the CRDM mechanism, where the energy peak of the DM flux increases with its mass. Interestingly, we find that in the framework of a dark photon mediator model, the sensitivities of  paleo-detectors obtained from the signals caused by the SN-sourced DM flux, even conservatively derived, can be five to ten orders of magnitude stronger than those by CR-boosted DM flux~\cite{Wang:2026you}.
For both mechanisms, we show that paleo-detectors with a 100 g$\cdot$Gyr exposure are able to significantly improve the   sensitivities to DM-nucleon scattering cross section for sub-GeV DM, exceeding current and near‑future direct detection experiments.

The rest of the paper is organized as follows. 
Sec.~\ref{sec:DM flux} presents the flux calculations for the CRDM mechanism with a constant spin‑independent cross section and for the SNDM mechanism with a dark photon mediator model.
Sec.~\ref{sec:events} describes the track length distribution from DM-induced nuclear recoils in paleo-detectors and summarizes all major background sources with their uncertainties.
Sec.~\ref{sec: Limits} presents the binned track length distribution and the statistical method, from which we derive 95\% C.L. sensitivity to the DM-nucleon scattering cross section for sub-GeV DM in both the CRDM and SNDM scenarios.
Sec.~\ref{sec:Conclusion} concludes the paper.

\section{Boosted DM Flux from both cosmic ray and supernova}
\label{sec:DM flux}

We start with the calculation of the boosted DM flux, which is either CR-boosted or SN-sourced.
In non-relativistic two-body kinematics, the maximum recoil energy transferred from a DM particle to a target nucleus at rest in elastic scattering is given by
\begin{equation}
T_N^{\text{max}} \approx \frac{2 m_\chi^2 v^2}{m_N}\,,    
\end{equation}
where $m_\chi $ ($m_N$) is the DM (target nucleus) mass, and $v$ is the incident DM velocity, typically around $10^{-3}c$ in our Milky Way. 
For sub-GeV DM colliding with the heavy target nuclei with mass $m_N$ between $10\gev$ and $100~\gev$, 
the recoil energy lies far below the keV  threshold of both current direct detection experiments and paleo-detectors. 
If DM particles are accelerated to (semi-)relativistic energies through processes such as CR-DM scattering or SN explosions, 
their kinetic energies $T_\chi$ can then exceed $(10\,\text{MeV})^2/m_
\chi$, 
thereby ensuring that the detectable signals lie in the range above the threshold of relevant nuclei-recoil experiments. 

\subsection{Cosmic-Ray–Boosted MeV DM Flux}
\label{sec:CRDM flux}
In the CRDM mechanism,  high-energy CR particles can transfer kinetic energy to DM particles via (in-)elastic scattering, accelerating the latter to (semi-)relativistic speeds~\cite{Cappiello:2018hsu,Bringmann:2018cvk}. 
The kinetic energy transferred to a DM particle in a single collision is given by
\begin{equation}
    T_\chi = \frac{1-\cos\theta}{2} T_\chi^{\max}, \quad ~{\rm with}~
T_\chi^{\max} = \frac{T_i^2 + 2m_i T_i}{T_i + \dfrac{(m_i+m_\chi)^2}{2m_\chi}}\,, 
\label{eq:T_chi_max}
\end{equation}
where $m_i$ and $T_i$ are the mass and kinetic energy of the incoming CR particle, $\theta$ is the scattering angle in the center-of-mass frame, 
and $T_\chi^{\max}$ corresponds to the maximal energy transfer in the case of back-to-back scatterings ($\theta = \pi$). 
Such (semi-)relativistic DM can then produce measurable nuclear recoil tracks in paleo-detectors, enabling the detection of MeV-mass DM particles. 

To quantize the sensitivities of relevant experiments, one first needs to calculate the total DM flux boosted by scatterings with CR: 
\begin{equation}
\frac{d\Phi_\chi}{dT_\chi} = D_{\mathrm{eff}} \frac{\rho^{\mathrm{local}}_\chi}{m_\chi} \sum_i \int_{T_i^{\min}}^\infty dT_i \, \frac{d\Phi^{\mathrm{LIS}}_i}{dT_i} \frac{\rm \sigma^0_{\chi i}}{T_\chi^{\rm max}(T_i)}F_i^2(q^2) \,,
\quad ~{\rm with}~\sigma_{\chi i}^0 = A_i^2 \sigma^{\rm SI}_{\chi p} \,\left[ \frac{m_N (m_\chi + m_p)}{m_p (m_\chi + m_N)} \right]^2\, ,
\end{equation}
where $\sigma^{\rm SI}_{\chi p}$ is the elastic scattering cross section in zero momentum transfer, and $F_i(q^2)$ is the nuclear form factor accounting for the suppression of coherent scattering at large momentum transfer, with $m_i$ ($A_i$) being the atomic mass (number) of the nucleus $i$. 
We adopt the effective distance $D_{\mathrm{eff}}=8 \kpc$~\cite{Bringmann:2018cvk}, 
\footnote{{Ref.~\cite{Xia:2021vbz} shows that $D_{\rm eff}$ can be energy-dependent, 
but the variation is only within $6$--$9~{\rm kpc}$ and has negligible impact on our results. 
}}
corresponding to a line-of-sight integration to $10 \kpc$ assuming a homogeneous CR distribution and an NFW DM profile~\cite{Navarro_1996,Ackermann_2012}, where 
the local DM density is taken as $\rho_\chi = 0.4\,\mathrm{GeV/cm^3}$. 
The lower integration limit $T_i^{\mathrm{min}}$ is the minimum kinetic energy of a CR particle 
required to accelerate a DM particle to reach a final kinetic energy $T_\chi$, given by 
\begin{equation}
T_{i}^{\min} = (\frac{T_\chi}{2}-m_i) \left[1 \pm \sqrt{ 1 + \frac{2 T_\chi (m_i + m_\chi)^2}{m_\chi( 2 m_i-T_\chi)^2} }\right],
\end{equation}
where the $+$ sign applies for $T_\chi > 2m_i$ and the $-$ sign for $T_\chi < 2m_i$.
Regarding the CR flux in our Milky Way, the local differential CR flux is fitted with 
\begin{equation}
\frac{d\Phi^{\mathrm{LIS}}_i}{dE_i} = \sum_{j} c_{ij} E_i^{-\alpha_{ij}} \exp\left( -\frac{E_i}{Z_i R_j} \right),
\end{equation}
where the values of the coefficients $c_{ij},~\alpha_{ij},~Z_i$, and $R_j$ are adopted from Ref.~\cite{Gaisser:2013bla,Xia:2020apm}. 
For the energy range relevant here, the CR spectrum is dominated by proton and $^4\mathrm{He}$, while the spectrum is significantly suppressed at higher energies, at PeV-scale and beyond.
To take into account the inner structure of CR particles, we follow~\cite{Bringmann:2018cvk,Perdrisat:2006hj} and adopt  for the proton the dipole form factor
\begin{equation}
    F_p(q^2) = \frac{1}{\left(1 + \left|q^2\right| / \Lambda_p^2\right)^2}\,,
\end{equation}
where $\Lambda_p= 770\ \text{MeV}$ is the cutoff scale for the proton. For all heavier nuclei, we instead use the Helm form factor~\cite{Duda:2006uk, Dent:2019krz}
\begin{equation}\label{eq:Helm}
    F_i(q^2) = \frac{3 j_1(q r_{i})}{q r_{i}} \exp({-|q^2| s^2 \over 2 })\,,
\end{equation}
where $r_{i} \approx 1.2\,A_i^{1/3}\ \text{fm}$ is the nuclear radius of species $i$, the surface thickness $s \approx 0.9\ \text{fm}$, and $j_1$ is the spherical Bessel function of the first kind.

\begin{table}[t]
\centering
\begin{tabular}{lccc}
\hline
Element &  $f_N$ (\%) &  $m_N$ (GeV) &  $A_N$ \\
\hline
Fe   & 4.8  & 52.1   & 56 \\
Si & 28.9 & 26.1  & 28 \\
O   & 46.5 & 14.9   & 16 \\
Al & 8.3  & 25.1 & 27 \\
\hline
\end{tabular}
\caption{Elemental composition and nuclear parameters of Earth  crust adopted, following Ref.~\cite{Zaharijas:2004jv}.}
\label{tab:crust_composition}
\end{table}

We also include the attenuation effect induced by Earth, as the DM particles must traverse a certain distance through Earth, taken to be 5\,km in this work, before reaching the paleo-detectors. During their propagation, DM particles gradually lose kinetic energy through scatterings with crustal elements of Earth.  The corresponding energy loss rate can be described by 
\begin{equation}\label{eq:CRDMDD}
\begin{aligned}
    \frac{dT_\chi}{dx} &= -\sum_N n_N \int_0^{T^{\text{max}}_r} \frac{d\sigma_{\chi N}}{dT_r} \, T_r \, dT_r =-\frac{1}{2} \sum_N n_N \sigma_{\chi N} T^{\text{max}}_r\,.\\
\end{aligned}
\end{equation}
Here we adopt the constant differential cross section approximation $d\sigma_{\chi N}/dT_r \simeq \sigma_{\chi N}/T^{\text{max}}_r$, 
while $T^{\text{max}}_r$, the maximal energy loss of the DM particle scattered off an at-rest nucleon $N$, has the same expression as  Eq.~\eqref{eq:T_chi_max} with the substitution $\chi \to N$ and $i\to \chi$. It is worth mentioning that constructing DM models with constant differential cross sections can be highly nontrivial for the CRDM mechanism~\cite{Ema:2020ulo}. We nonetheless adopt this approximation to allow a direct comparison between the sensitivities of paleo-detectors and the limits set by previous studies.
The nuclear number density $n_N=\rho_{\text{Earth}}f_N/m_N$ for each element in   Earth crust is derived from the Earth density model~\cite{Zaharijas:2004jv,Emken:2017qmp}, 
where  
$f_N$ is the mass fraction of the element in the crust, and 
$m_N$ is the nuclear mass. The crustal density is given by $\rho_{\text{Earth}}= 2.6\ \text{g/cm}^3$, and the dominant crustal elements and their parameters are summarized in Table~\ref{tab:crust_composition}.

\begin{figure*}[t]
\centering
\includegraphics[width=8.6cm]{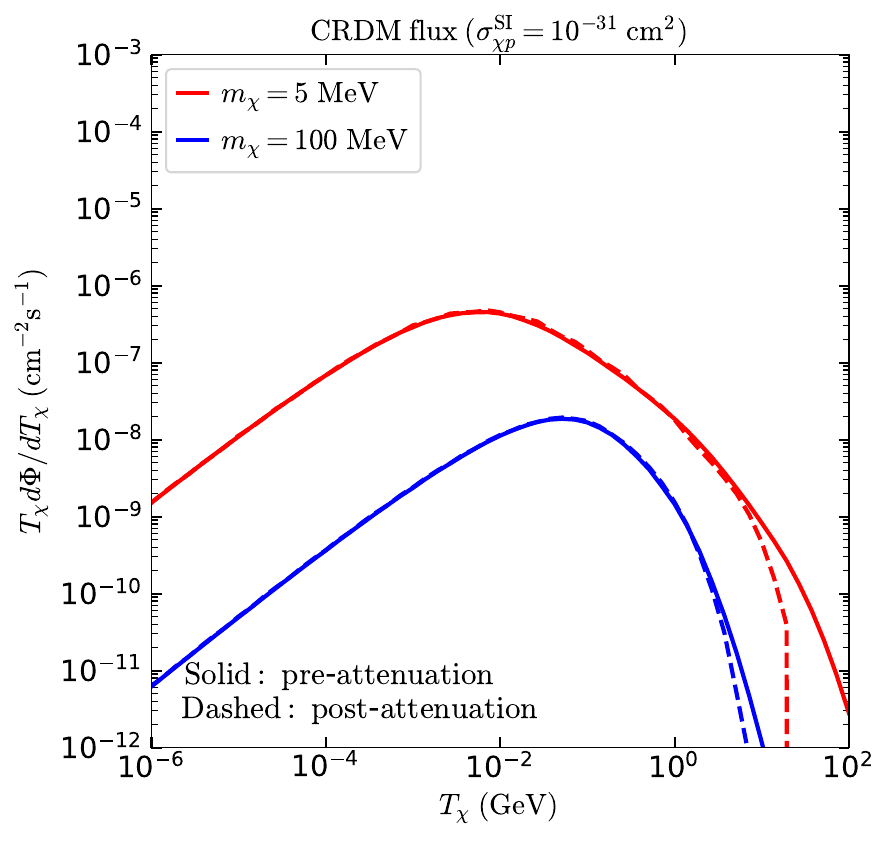}~
\includegraphics[width=8.6cm]{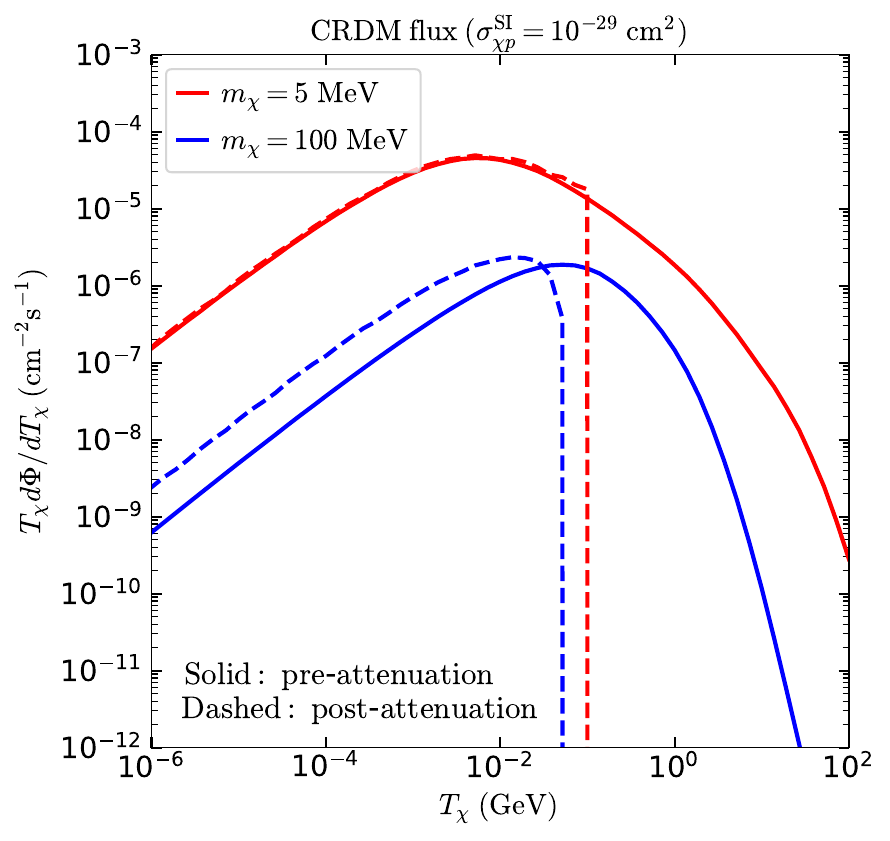}
 \caption{The energy spectrum of CR–boosted DM flux for various DM masses, with a constant DM-nucleon cross section $\sigma^{\text{SI}}_{\chi p} = 10^{-31}~\text{cm}^2$ (left panel) and  $\sigma^{\text{SI}}_{\chi p} = 10^{-29}~\text{cm}^2$ (right panel). 
 The red and blue lines correspond to DM particle masses of $m_\chi = 5~\mev$ and $100~\mev$, respectively. The solid lines show the initial cosmic ray boosted DM flux, while the dashed lines illustrate the corresponding flux after including the attenuation induced by Earth (setting the depth $z=5$\,km). 
 }
 \label{fig:CRDM_flux}
 \end{figure*}

The CR-boosted DM flux, before and after being attenuated by scattering over about 5\,km of Earth’s crust, is illustrated in Fig.~\ref{fig:CRDM_flux} for two constant cross sections,  $\sigma^{\text{SI}}_{\chi p} = 10^{-31}~\text{cm}^2$ (left panel) and $10^{-29}~\text{cm}^2$ (right panel).
The solid lines represent the CR boosted DM flux before attenuation for two different masses, $m_\chi = 5~\mev$ (red) and $m_\chi = 100~\mev$ (blue). 
It shows that the flux of lighter DM exceeds that of heavier DM, mainly due to the fact that given the fixed local DM mass density, its number density scales inversely with its mass. 
Another important effect induced by the DM mass is that for the same kinetic energy of incoming CR particle $T_i$, heavier DM can be boosted to obtain larger kinetic energy, with larger $T^{\rm max}_
\chi$. Consequently, the boosted flux of a heavier DM at the low $T_\chi$ end is further suppressed if the total DM-nucleon scattering cross section is fixed. 
In other words, it makes the boosted DM energy spectrum peak at higher energies for heavier DM. This explains the difference of the peak locations in Fig.~\ref{fig:CRDM_flux}, and, together with the number density argument above, yields the ${\mathcal O}(10^2)$  suppression of DM flux for $m_\chi =100\,$MeV (solid blue line) with respect to that for for $m_\chi =100\,$MeV (solid red line)  at the low-$T_\chi$ end.

The dashed lines in two panels of Fig.~\ref{fig:CRDM_flux} indicate the energy spectra of the DM flux after accounting for attenuation effects of   Earth. As the kinetic energy of DM increases, the energy loss per unit distance becomes more important, 
leading to a reduction in the flux of high-energy DM particles. 
Consequently, in the high-energy region, the spectrum after attenuation (dashed line) shifts toward lower energies compared to  the   spectrum of pre-attenuation (solid line). 
In the low-energy region, where the attenuation effect is negligible, the spectrum remains nearly unchanged.
Besides,  the right panel of  Fig.~\ref{fig:CRDM_flux} explicitly shows that   Earth attenuation becomes significant for $\sigma^{\rm SI}_{\chi p} \gtrsim 10^{-29}\,\text{cm}^2$, thus we set this value as the maximum of the $y$‑axis when generating the sensitivity plot for the CRDM mechanism below.

\subsection{Supernova-Sourced MeV DM Flux}
\label{sec:SNDM flux}

Now we turn to the SNDM mechanism, in which an energetic DM flux is supposed to be generated from the hot dense core of nearby SN events, as long as the DM mass is not too larger than the core temperature~\cite{DeRocco:2019jti}. In this case, one needs to obtain both the production cross section of the DM particles and the scattering cross section between DM and nucleus targets, for which a concrete particle model is preferred. Thus, following \cite{DeRocco:2019jti, Wang:2026you} 
we consider a dark photon model with the interaction Lagrangian
$$
\mathcal{L}_{\rm int} = g_\chi (\bar{\chi} \gamma^\mu \chi)  A'_\mu + \epsilon e J^\mu_\text{em} A'_\mu\,,
$$
where $\chi$ and $A^{\prime}$ denote the fields of DM particles and the  dark photon mediator, respectively. In addition, $g_\chi$ and $e$ ares the dark and electromagnetic gauge couplings, as the dark photon kinetically mixes with the photon via a tiny dimensionless parameter $\epsilon$. That is, the interaction between dark photon and  nucleus with charge number $Z$ is  characterized by  $\epsilon (Z e)$. 
We also introduce a dimensionless coupling 
\begin{equation}
    y \equiv  \epsilon^2 {g_\chi^2 \over 4\pi} \left( m_\chi \over m_{A'}\right)^4 \simeq   10^{-21} \left({\epsilon \over 10^{-6} }\right)^2 \left({g_\chi\over 0.3 }\right)^2 \left({m_\chi  \over 20\,\text{MeV} }\right)^4  \left({\text{GeV} \over m_{A'}}\right)^4 \,
\end{equation} 
 to parametrize the effective strength of the portal interaction, where $m_{A'}$ is the dark photon mass.

With the kinetic mixing between the photon and dark photon, sub-GeV DM can be produced in the SN core via electron-positron annihilation ($e^+e^- \leftrightarrow \bar{\chi}\chi$) and proton-neutron bremsstrahlung ($np \leftrightarrow np\bar{\chi}\chi$), among other processes. 
Once produced, DM should escape from the core  freely for small enough portal interactions, which is usually referred to as the free-streaming regime. For larger values of $y$ above this regime, produced DM particles are partially or fully captured, via scattering with protons and electrons inside the SN core, which is the trapping regime. This happens below the so-called decoupling surface of the SN core, whose radius increases with larger $y$.  For both regimes, one can approximate the  energy spectrum of the emitted DM flux with a Fermi-Dirac distribution~\cite{DeRocco:2019jti, Bhalla:2025vnq}, whose differential flux can be written as
\begin{equation}
\frac{d\Phi_{\chi,\rm{SN}}}{dE}(m_\chi, y) = \Phi_\chi (m_\chi, y)
\left( \frac{E^2 - m_\chi^2}{\exp(E/T) + 1} \right)
\left( \int_{m_\chi}^{\infty} \frac{E^2 - m_\chi^2}{\exp(E/T) + 1} dE \right)^{-1}\,,   
\end{equation}
where $T$ is either the averaged SN temperature in the free-streaming regime, or the temperature of the decoupling surface in the trapping regime. Following Ref.~\cite{Bhalla:2025vnq}, we always set $T = 30$ MeV for simplicity. 
In turn, the flux of the SN-sourced DM particles is obtained by integrating over the spatial distribution of SN events within the Galactic halo in the past via
\begin{equation}
\Phi_\chi(m_\chi, y) = N_\chi(m_\chi, y) \int_{0}^{\pi} \sin \theta_s \, d\theta_s \int_{0}^{2\pi} d\phi_s  \int_{0}^{l_s^{\text{max}}}
\frac{1}{4\pi l_s^2} \,
\frac{dN_{\text{SN}}}{dt}(r,z) \,
l_s^2 \, dl_s \,,
\end{equation}
where  $l_s$ denotes the distance from each location to Earth, whose integration upper limit is set as $l_s^{\text{max}} = 10$ kpc, which suffices to cover most of the Galactic disk. In addition,  $dN_{\text{SN}}/dt$ denotes the SN rate density as a function of its possible location in the Galactic cylindrical coordinates $(r, z)$, estimated by an exponential function~\cite{Adams:2013ana} as
\begin{equation}
\frac{dN_{\text{SN}}}{dt}(r, z) = 2.08\times 10^{-3} \, \text{kpc}^{-3} \text{yr}^{-1}\times \exp(-{r \over 2.9 \,\text{kpc}})\, \exp(-{|z| \over 95 \,\text{pc}}) \,    \,.
\end{equation}
The transformation from the cylindrical coordinates $(r, z)$ to Earth-centric spherical coordinates $(l_s, \theta_s, \phi_s)$ is given by
$
r= (r_E^2 + l_s^2 - l_s^2 \sin^2 \theta_s \sin^2 \phi_s - 2 l_s r_E \cos \theta_s )^{1/2}$ and  $
z = l_s \sin \theta_s \sin \phi_s,
$ where our Earth is located at $(r, z) = (r_E, z_E)$ with $r_E = 8.7$\,kpc and $z_E = 24$\,pc. 
At last, the pre-factor $N_\chi$ yields the averaged total number of DM particles produced per single SN event. 

\begin{figure*}[t]
\centering
\includegraphics[width=8.3cm]{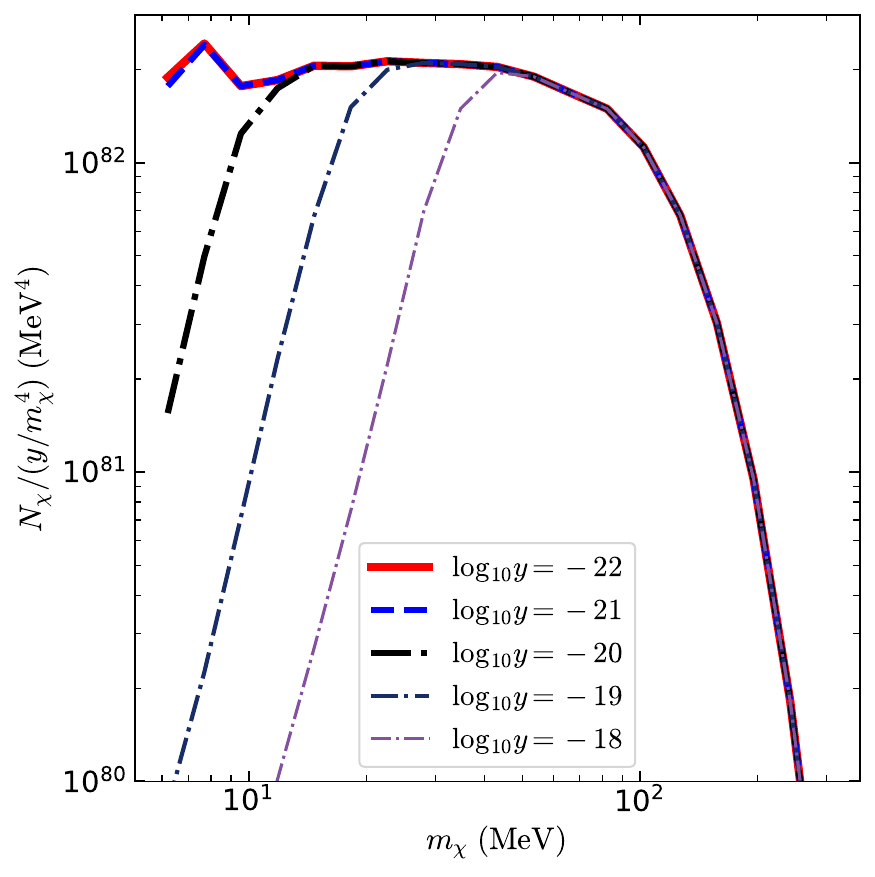}~~
\includegraphics[width=9cm]{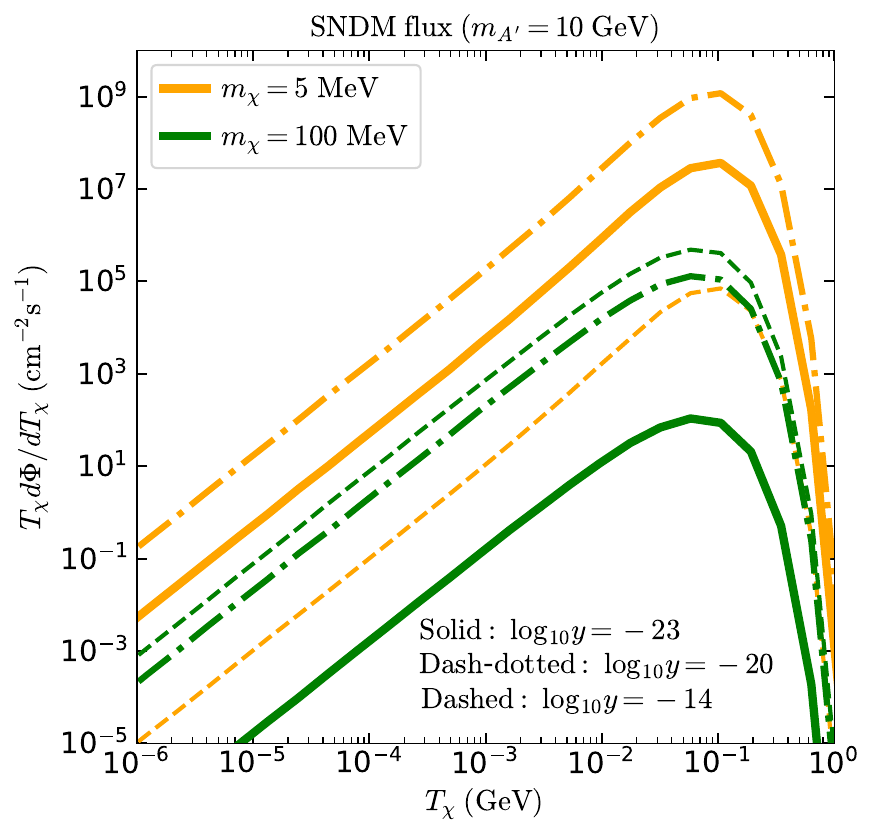}
 \caption{{\bf Left panel:} Representative values of SN-sourced DM numbers $N_\chi$ for various effective couplings $y$ and DM mass $m_\chi$, fitted using the data table provided in  Ref.~\cite{DeRocco:2019jti}. The values have been normalized with $y/m_\chi^4$, which is  proportional to the DM production cross section at relativistic limit.  {\bf Right panel:}  SN-sourced DM fluxes for $m_\chi = 5 \mev$ (orange) and $100 \mev$ (green). In both panels, the $y$  values are chosen such that, with increasing $y$, the flux first increases (in the free-streaming regime) and then decreases (in the trapping regime); see the main text for detailed discussions. }
 \label{fig:SNDM_flux}
 \end{figure*}

The value of $N_\chi$ depends on the effective coupling $y$ and the DM mass $m_\chi$, and needs to be estimated from numerical simulations in the trapping regime. Here we directly subtract its values for the parameter region of our interest from the data table provided in  Ref.~\cite{DeRocco:2019jti}.  Some representative numbers are presented in left panel of Fig.~\ref{fig:SNDM_flux}.  In the free-streaming regime, DM particles produced in the proto-neutron star escape almost freely, leading to the scaling of  $N_\chi$ linearly as  $\epsilon^2 g_\chi^2 / (4 \pi m_{A'}^4)$,  or equivalently, as $y /m_\chi^4$. This is demonstrated by the overlap of the red solid ($y=10^{-22}$) and blue dashed ($y=10^{-21}$) lines in left panel of Fig.~\ref{fig:SNDM_flux}. 
As $y$ increases, the scattering cross section between DM and nucleons becomes sizable, leading to its efficient trapping within the SN core. In this trapping regime, the emitted DM flux starts to decrease with increasing values of $y$. Notably, for DM with masses well below the SN temperature, the trapping efficiency with protons is also proportional to  $y /m_\chi^2$, suggesting that lighter DM enters this trapped regime at smaller $y$ values. This is shown by the dot-dashed lines in the left panel of Fig.~\ref{fig:SNDM_flux}, where an effective coupling as large as $y=10^{-20}$ ($y=10^{-18}$) starts to suppress the flux for DM masses below 15\,MeV (40\,MeV) with respect to the expected values if DM free-streams.


Right panel of Fig.~\ref{fig:SNDM_flux} in turn  shows the SN-sourced DM flux at the location of Earth for different effective coupling $y$, with  DM masses $m_\chi = 5 \mev$ (orange lines) and $m_\chi = 100 \mev$ (green lines). First of all, the kinetic energy distribution of SN-sourced DM follows a Fermi-Dirac distribution, peaking at around 100\,MeV and falling off exponentially at higher energies.  
Secondly, it demonstrates the same dependence on $y$ and $m_\chi$ as we have observed from the left panel. 
Taking the lines for $m_\chi = 5 \mev$ (orange ones) as an example, the dependence of the SNDM flux on $y$ can be understood in the following way. As $y$ increases from a very tiny value, the flux first rises from $y=10^{-23}$ (orange solid) to $y=10^{-20}$ (orange dash-dotted), and then drops sharply, with $y=10^{-14}$  (orange dashed) lying nearly three orders of magnitude below the solid one. This behavior reflects the competition between DM production and trapping. At small $y$, DM escapes freely and $N_\chi$ grows linearly with $y$, while at larger $y$, DM-proton/electron scattering becomes frequent enough to trap DM inside the protoneutron star, suppressing the emitted DM  flux. 
The $m_\chi = 5 \mev$  (orange) and  $m_\chi = 100 \mev$ (green) lines directly compare the SNDM flux for different DM masses.
At very small couplings (thick solid) in the free-streaming regime, lighter DM yields a larger flux, with the $m_\chi = 5 \mev$ case exceeding the $m_\chi = 100 \mev$ one by about five orders of magnitude, as expected from the flux dependence on $y/m_{\chi}^4$. At very large couplings (thin dashed) in the trapping regime, the trend reverses because a smaller DM mass leads to a larger $\sigma_{\chi p}$ for a fixed value of $y$, making lighter DM particles more efficiently trapped and thus its flux more strongly suppressed. 

\section{Track Length Distributions}
\label{sec:events}
\subsection{SIGNALS}
Ancient minerals as DM detectors accumulate traces of nuclear recoils over geological timescales (hundreds of millions of years), offering an effective exposure of hundreds of ton-years in a cost-effective manner. 
 We focus on olivine $\rm (Fe, Mg)_2SiO_4$ as the target material in this work.

DM particles scatter off target nuclei within a mineral, producing primary recoil atoms. 
These recoil atoms subsequently migrate through the crystal lattice and create damage tracks that can be identified with microscopy techniques~\cite{Lider_2017, HILL201265}. 
The associated differential nuclear recoil rate per unit target mass is  expressed as  
\begin{equation}
    \frac{dR}{dE_R}(E_R)=\frac{1}{m_T}\int dT_\chi \frac{d\Phi_{\chi}}{dT_\chi} \frac{d\sigma}{d E_{R}}(E_R)\,,
\end{equation}
where $m_T$ is the mass of the target nucleus, and $E_R$ is the recoil energy. For a constant scattering cross section, referred to as $\sigma^{\rm SI}_{\chi p}$ below, adopted in the CRDM mechanism, its differential form is the same as ${d\sigma}/{d T_{r}}$ introduced in Eq.~\eqref{eq:CRDMDD} by replacing $T_r$ with $E_R$. For the dark photon model used for the SNDM mechanism,  the differential DM-nucleus scattering cross section, up to its electromagnetic form factor, is approximately 
\begin{equation}
\frac{{\rm d}\sigma}{{\rm d}E_R} \simeq \frac{Z^2 \sigma_{\chi p}}{\mu_p^2} \frac{2m_T (m_\chi +T_\chi)^2 - E_R m_T^2}{4  T_\chi (2m_\chi +T_\chi)}\,,
\end{equation}
where $m_T$ is again the target nucleus mass with charge $Z$, and the DM-nucleon reduced mass $\mu_p = m_\chi m_p/(m_\chi+m_p)$. We have used the approximation that $\max[T_\chi , m_\chi]\ll m_T$. With this approximation, there is $E_R^{\rm max} \simeq 2T_\chi (2m_\chi + T_\chi)/m_T$.  For comparison with the CRDM results of both ours and Ref.~\cite{Wang:2026you}, we have defined a constant (energy-independent) counterpart in the form of DM-proton scattering cross section 
\begin{equation}\label{eq:effchip}
\sigma_{\chi p} \equiv   \frac{\epsilon^2 e^2 g_\chi^2 }{\pi}\frac{ \mu_p^2}{ m_{A'}^4} \simeq 10^{-44}\,{\rm cm}^2 \left({\epsilon \over 10^{-6} }\right)^2 \left({g_\chi\over 0.3 }\right)^2 \left({\mu_p  \over 100\,\text{MeV} }\right)^2  \left({\text{GeV} \over m_{A'}}\right)^4 \,.
\end{equation}
For more detailed calculation of relevant direct detection cross sections, see \emph{e.g.} Ref.~\cite{Bringmann:2018cvk,Perdrisat:2006hj, Duda:2006uk, Dent:2019krz}.

\begin{figure*}[t]
\centering
\includegraphics[width=8.3cm]{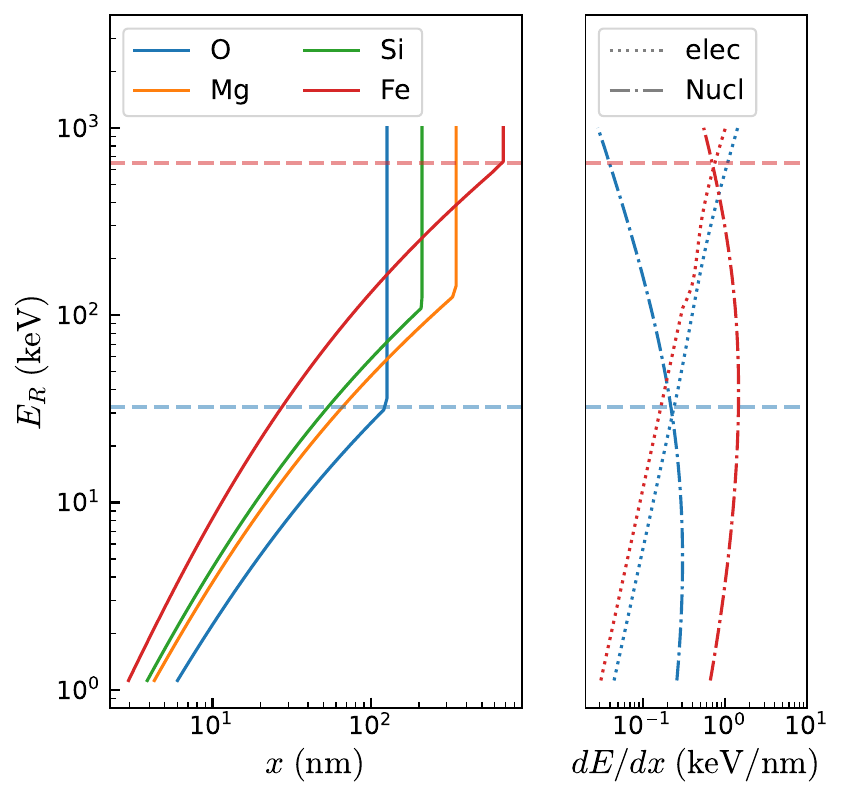}~~
\includegraphics[width=8.6cm]{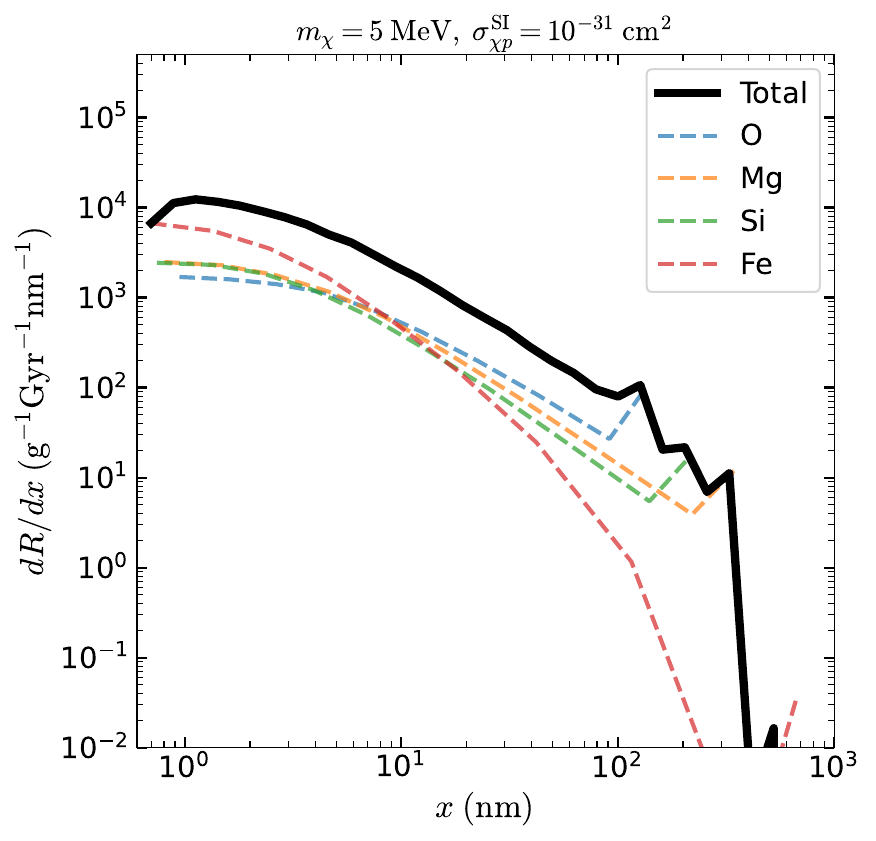}
 \caption{
\textbf{Left panel}: Track length and the stopping power as a function of nuclear recoil energy $E_R$. The solid lines in blue, green, orange, and red correspond to oxygen (O), silicon (Si), magnesium (Mg), and iron (Fe) nucleus in olivine molecules, respectively, illustrating the $E_R-x$ relation. The dash-dotted and dotted lines show the nuclear and electronic stopping power for O and Fe, respectively. We manually fix the length of tracks, when the electronic stopping dominates over that of nucleus, as indicated by the two horizontal dashed lines. 
\textbf{Right panel}: Differential nuclear recoil track distribution of the signals in the case of CRDM. The colored dashed lines represent the contributions of various nuclei, while the black solid line shows the total differential event rate. The results are shown for a benchmark parameter choice with $m_\chi = 5$ MeV and $\sigma_{\rm SI} = 10^{-31}$ cm$^2$.
 }
 \label{fig:dR_dx}
 \end{figure*}

Using the differential nuclear recoil rates given above, now we calculate the track length distribution with 
\begin{equation}\label{eq:dRdx}
\frac{dR}{dx}(x) = \sum_i \frac{dR_i}{dE_R}(E_R) \cdot \left( \frac{dE}{dx}(E_R) \right)_i\, ,
\end{equation}
where $i$ indexes the different target elements in the mineral, and $({dE}/{dx})_i$ is the corresponding nuclear stopping power for that element. The latter describes the energy lost by a recoiling nucleus per unit path length as it travels through the ancient mineral, and is calculated with the \texttt{SRIM} software~\cite{ZIEGLER20101818}.
Concretely, the expected average track length for a recoil energy  can be obtained via 
\begin{equation}
x^i_{\mathrm{avg}}(E_R) = \int_0^{E_R} \left( \frac{dE}{dx}(E) \right)_i^{-1} dE\, ,
\end{equation}
whose derivative in turn yields the value of $({dE}/{dx})_i$, as a function of $E_R$, used in Eq.~\eqref{eq:dRdx}. The corresponding numbers are presented in the left panel of Fig.~\ref{fig:dR_dx}, showing the track length and the stopping power as functions of the initial recoil energy $E_R$.   
The color of the lines indicates different elements, \emph{i.e.}, oxygen (O in blue), silicon (Si in green), magnesium (Mg in orange), and iron (Fe in red), respectively, while, for further demonstration, the dash‑dotted and dotted lines denote the nuclear and electronic stopping power for oxygen (O) and iron (Fe).
The recoiling nucleus transfers kinetic energy through elastic collisions,  displacing atoms and forming damage tracks. 

In the low recoil-energy region with $E_R \lesssim 20$-$100\kev$, the nuclear stopping power dominates the energy loss of the recoiling nucleus. As the recoil energy increases, the nucleus travels farther before coming to rest, resulting in a track length that increases with \(E_{\rm R}\), as illustrated in the left panel of Fig.~\ref{fig:dR_dx} for  the regime \(x \lesssim 10^{2}\)\,nm. 
At low recoil energies, Fe yields the shortest track length for a given value of $E_R$,  because its large nuclear stopping power causes it to lose energy rapidly per unit path length. In other words, lighter nuclei with the same initial kinetic energy are able to travel longer in a paleo-detector.  
Beyond a certain energy threshold, electronic stopping power becomes the dominant energy-loss mechanism of the recoiling nucleus, indicated by the dashed horizontal lines for O and Fe. In this case, the recoiling nucleus deposits most of its kinetic energy into exciting and ionizing electrons rather than displacing atoms. The kinetic energy is thus dissipated over a short distance through ionization, severely limiting atomic displacements. As a result, the track length stop increasing with recoil energy and instead reaches a truncation value beyond the threshold. For heavier recoils, which possess a higher nuclear stopping power, this truncation length shifts to larger values. In this plot and our numerical implementation, we impose this length truncation manually once electronic stopping becomes dominant.


With the differential scattering cross section and the $E_R$-$x$ relation, we draw in the right panel of Fig.~\ref{fig:dR_dx} the differential nuclear recoil track distribution of the CRDM scenario as a function of nuclear recoil energy. Given a fixed  kinetic energy $T_\chi$ of sub-GeV DM, heavier nuclei would receive smaller recoil energies after scattering, and thus result in shorter recoil tracks.  While the number of DM particles at the high‑energy end is smoothly suppressed, the track length at the long end is truncated due to the dominance of electron stopping.  The SNDM case is expected to exhibit similar features, while with its kinetic energy peaking around hundreds of MeV, its track length distribution tends to have a more pronounced long-track tail.

\subsection{BACKGROUNDS}
We adopt the background model of paleo-detectors that has been developed in Refs.~\cite{Drukier:2018pdy,Baum:2021jak}. This model provides detailed spectra and event rates for all relevant background sources in paleo-detection. Each component is briefly summarized as follows: 
\begin{itemize}
\item \textbf{Solar neutrinos}~\cite{Gonzalo:2023mdh,Gonzalez-Garcia:2023kva}: They dominate the irreducible neutrino background and contribute primarily to track lengths $x < 100 \; \rm nm$. 
We assign a conservative 100\% normalization uncertainty to account for possible time-averaged flux variations over the Gyr exposure of paleo-detectors, denoted as $\sigma_{\text{solar}\; \nu}$, corresponding to the theoretical error on the $^8$B neutrino flux~\cite{Serenelli:2016nms}.

\item \textbf{SN neutrinos}~\cite{Baum:2019fqm}: We include both the Diffuse SN Neutrino Background (DSNB) and the Galactic SN Neutrino Background (GSNB). 
Their fluxes are taken from Ref.~\cite{Baum:2019fqm}, with a conservative 100\% theoretical uncertainty~\cite{Fung:2025cub}. 
We denote the nuisance parameters for the DSNB and GSNB normalizations as $\sigma_{\text{DSNB}}$ and $\sigma_{\text{GSNB}}$, respectively, 
and treat them independently.

\item \textbf{Atmospheric neutrinos}~\cite{Baum:2021jak,OHare:2020lva}: Their flux is subdominant but contributes to tracks beyond $100 \; \rm nm$. 
We adopt the fluxes from Ref~\cite{OHare:2020lva} and conservatively assign a 100\% uncertainty~\cite{Fung:2025cub}, denoted as $\sigma_{\text{atm}}$.

\item \textbf{Neutrons from $^{238}$U decay chains}~\cite{Baum:2018tfw,Drukier:2018pdy,Baum:2019fqm,Baum:2021jak}: This is typically the dominant background in paleo-detection. The neutron flux scales linearly with the $^{238}$U concentration. Based on  inductively coupled plasma
mass spectrometry measurements, we adopt a benchmark concentration of $10^{-10} g/g$ with a 1\% uncertainty~\cite{Fung:2025cub}, 
denoted as $\sigma_{\text{U}}$. 
The spectral shape of nuclear recoils from neutrons is taken from \texttt{paleoSpec}~\cite{Baum:2018tfw,Drukier:2018pdy,Baum:2019fqm,Baum:2021jak, KONING20122841, WILSON2009608,Soppera:2014zsj}.

\item \textbf{$^{234}$Th recoils}~\cite{Drukier:2018pdy}: The first $\alpha$ decay in the $^{238}$U chain produces a monoenergetic $^{234}$Th nucleus with $72 \kev$ recoil energy. The number density scales linearly with the $^{238}$U concentration. For our benchmark values, the $^{234}$Th density is $\sim 10^{7}\; g^{-1}$~\cite{Drukier:2018pdy}. Its uncertainty is inherited from $\sigma_{\text{U}}$ above.

\item \textbf{Intrinsic crystal defects}~\cite{ejm-33-249-2021}: Linear dislocations can mimic nuclear recoil tracks but typically extend over grain scales $\gtrsim 10^{3} \; \rm nm$, making them distinguishable from the $\mathcal{O}(1)$-$\mathcal{O}(10^2) \rm \; nm$ signals of interest.  Given the large background uncertainties of  such long tracks, we always restrict our analysis to tracks below $10^{3} \; \rm nm$. This is a much more conservative approach, being different from Ref.~\cite{Wang:2026you}. Thus we do not assign a nuisance parameter to this background. 
\end{itemize}

\begin{figure*}[t]
\centering
\includegraphics[width=8.6cm]{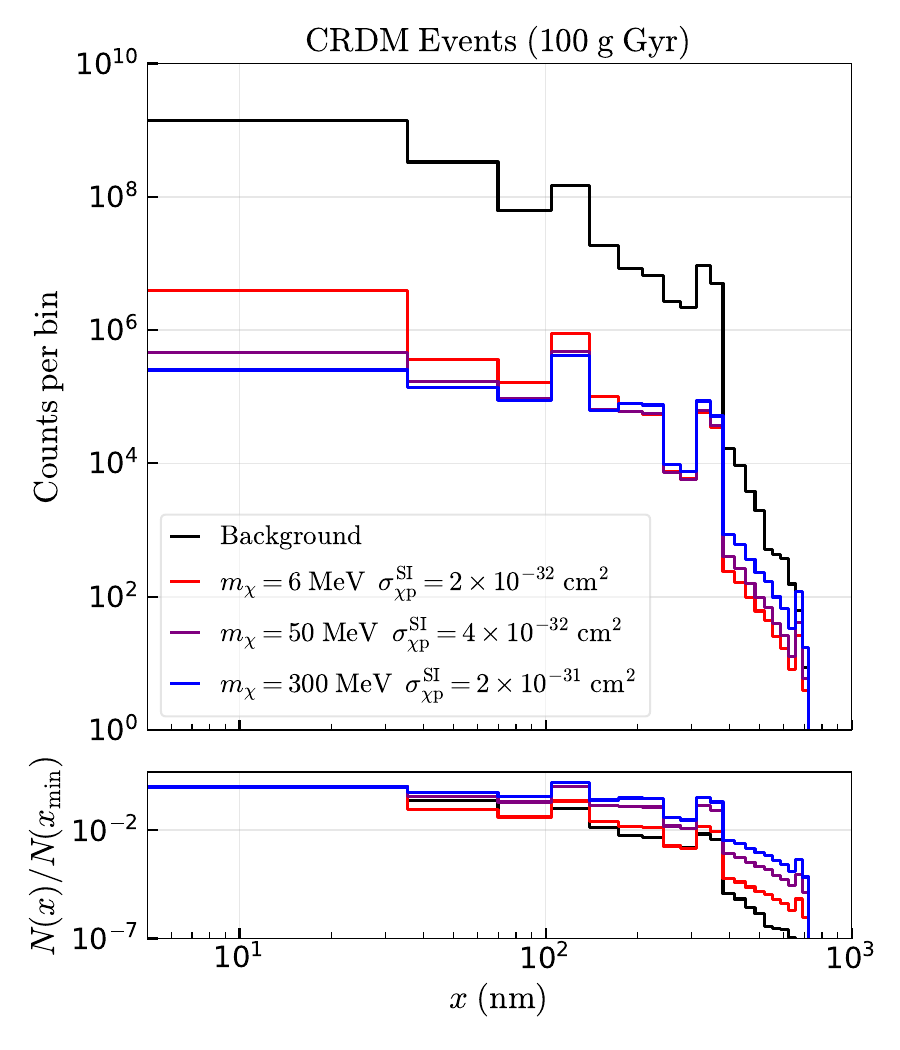}~~
\includegraphics[width=8.6cm]{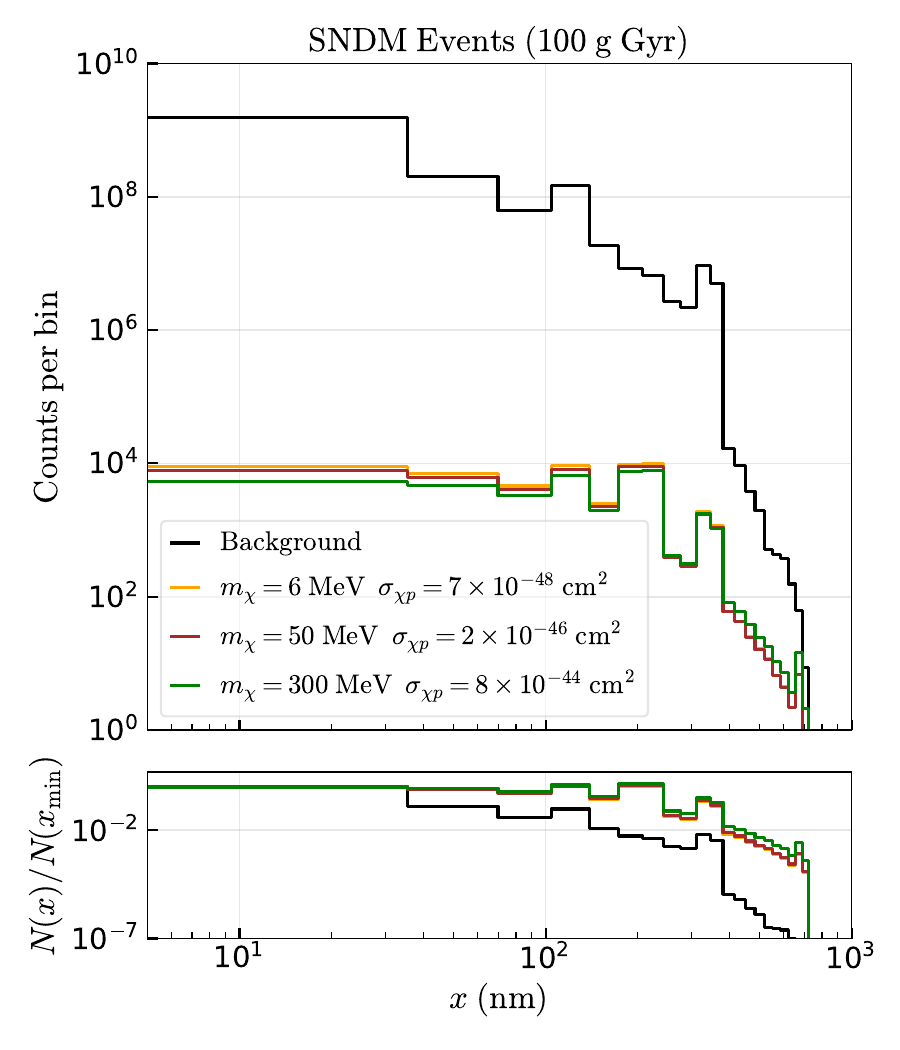}
 \caption{Binned Track Length Distributions for boosted DM. The black line represents the background track distribution. The upper panels show the
binned event counts, while the lower panels show the normalized
distributions $N(x)/N(x_{\rm min})$. The normalized distributions provide
a direct comparison of the track-length spectral shapes of the signals
and backgrounds. \textbf{Left panel}: Nuclear recoil track distribution event counts for CRDM. The red, purple, and blue lines correspond to $m_\chi = 6 \mev$, $m_\chi = 50 \mev$, and $m_\chi = 300 \mev$, respectively. \textbf{Right panel}: Nuclear recoil track distribution event counts for SNDM. The orange, brown, and green lines correspond to the same DM masses as in the left panel.
 }\label{fig:binned events}
 \end{figure*}

In practice, we divide the track length distribution into $N_{\text{bins}} = 30$ bins, and present the resulting histogram in Fig.~\ref{fig:binned events} for CR-boosted DM (left panel) and SN-sourced DM (right panel) with several different masses. The colored lines represent the event counts for masses of $6 \mev$ (red), $50 \mev$ (purple), and $300 \mev$ (blue),  while the black line shows the  distribution of the total background.  {The lower panels show the normalized distributions
\(N(x)/N(x_{\rm min})\),  where \(x_{\min}\) is the minimum track length in
the binned spectrum.}
These distributions are used to compare the track-length spectral shapes
of the signals and backgrounds.
For CRDM, heavier DM produces more events with long track lengths, driven by both a flux peaked at higher energies and more efficient energy transfer to target nuclei. 
In contrast, SNDM exhibits nearly overlapping track length distributions across different masses, as its DM flux always peaks at the kinetic energy around $100$\,MeV, solely determined by the SN temperature. Moreover, the normalized distributions show that the boosted-DM signals have
harder long-track tails than the dominant background. Equivalently, the
signal spectra fall more slowly at large track lengths. This spectral
difference cannot be absorbed by a global rescaling of the background.
It is particularly visible in the region
$x\simeq 100$--$600~{\rm nm}$, where radiogenic neutrons from the
$^{238}{\rm U}$ decay chain dominate the background. The binned spectral
analysis below uses this shape information, rather than relying only on
the absolute signal-to-background ratio in a single bin.



\section{Sensitivity Projections}
\label{sec: Limits}

This section describes how we quantitatively derive the projected sensitivities to the CRDM and SNDM mechanisms, following previous studies on paleo-detectors. The theoretical track count, including both signals and backgrounds, in the 
$i$-th bin is given by
\begin{equation}
R_i = \int dx' \, \frac{dR}{dx}(x') \, W_i(x')\,,    
\end{equation}
where $W_i(x')$ is a window function that describes the resolution effect, in terms of
\begin{equation}
W_i(x') = \frac{1}{2} \left[ \operatorname{erf}\!\left( \frac{x' - x_i^{\text{L}}}{\sqrt{2}\,\sigma_x} \right) - \operatorname{erf}\!\left( \frac{x' - x_i^{\text{R}}}{\sqrt{2}\,\sigma_x} \right) \right],
\end{equation}
where $x_i^{\text{L}}$ and $x_i^{\text{R}}$ denote the left and right edges of the $i$-th bin, respectively, while $\sigma_x=10 \; \rm nm$ is the resolution. This formulation follows  Ref.~\cite{Baum:2021jak} and accounts for the bin migration due to the finite resolution.

We then construct the test statistic $\chi^2$ using a background-only Asimov dataset,  as described in Ref.~\cite{Fung:2025cub}  
\begin{equation}
\chi^2(\{\vartheta\};\theta_j) = \sum_{i=1}^{N_{\text{bins}}} \frac{(R_{\text{data},i} - R_{\text{th},i}(\{\vartheta\},\theta_j))^2}{R_{\text{th},i}(\{\vartheta\},\theta_j)} + \sum_{j} \frac{(\theta_j - \theta_{\text{central},j})^2}{\sigma_j^2}\,.
\end{equation}
The first term compares the theoretical track count with the Asimov reference count per bin. 
The second term adds Gaussian priors for the nuisance parameters that account for background uncertainties. 
The theoretical track count for the $i$-th bin,  $R_{\text{th},i}$, depends on the DM parameters, $\{\vartheta\}$, and on the nuisance parameters, $\theta_j$, which control the normalizations of the backgrounds. The corresponding count from a background-only Asimov dataset is denoted  as $R_{\text{data},i}$. The nuisance parameters are constrained by Gaussian priors with central values $\theta_{\text{central},j}$ and uncertainties $\sigma_j$,  where $j$ indexes the different background sources listed above (\emph{e.g.},  $\sigma_{\text{solar}\; \nu}$, $\sigma_{\text{GSNB}}$, $\sigma_{\text{U}}$).
We determine the experimental sensitivities using the difference in $\chi^2$  values:
\begin{equation}
\Delta\chi^2(\{\vartheta\}) = \chi^2(\vartheta;\hat{\hat{\theta}}) - \chi^2(\hat{\vartheta};\hat{\theta})\,,       
\end{equation}
where $\hat{\vartheta}$ and $\hat{\theta}$ are the global best-fit parameters that minimize $\chi^2$, while $\hat{\hat{\theta}}$ indicates the conditional best-fit nuisance parameters for a fixed set of $\{\vartheta\}$. Parameter choices that yield $\Delta\chi^2 > 2.71$ are supposed to be experimentally excluded at  95\% C.L.\,.

In our work, the \(\chi^2\) analysis does not use independent bin-by-bin systematic errors. Instead, Eq. (20) fits the full binned track-length spectrum with correlated normalization nuisance parameters. This keeps the spectral-shape information. 
Importantly, the boosted-DM sensitivity mainly comes from the long-track region \(x\simeq100\)--\(600~{\rm nm}\). In this region, the background is dominated by radiogenic neutrons from the \(^{238}{\rm U}\) decay chain. The leading uncertainty of this background is set by the uranium concentration in the sample and by the exposure, which do not depend on track length, 
thus their leading effect is to rescale the overall \(^{238}{\rm U}\) background normalization.
Although detector response and related effects may induce residual shape
uncertainties, absorbing the boosted-DM signal would require a strongly
bin-dependent distortion of the \(^{238}{\rm U}\) track-length spectrum,
with fractional shifts varying by orders of magnitude across the relevant
long-track bins.  {We also provide a conservative robustness check in the appendix~\ref{Robustness check with bin-wise background uncertainty}, using a bin-wise variance form for the $\chi^2$.} 
This treatment weakens the correlated spectral-shape information, and   the projected sensitivity becomes weaker by about a factor of two.



\begin{figure*}[t]
\centering
\includegraphics[width=8.85cm]{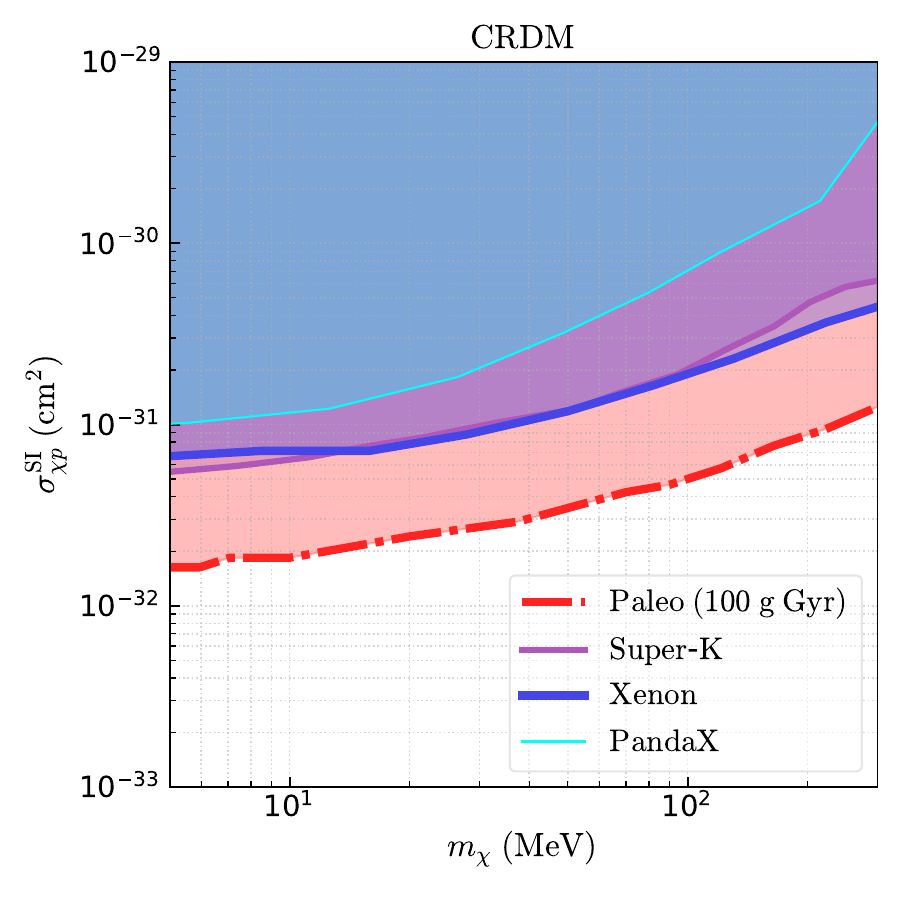}
\includegraphics[width=8.85cm]{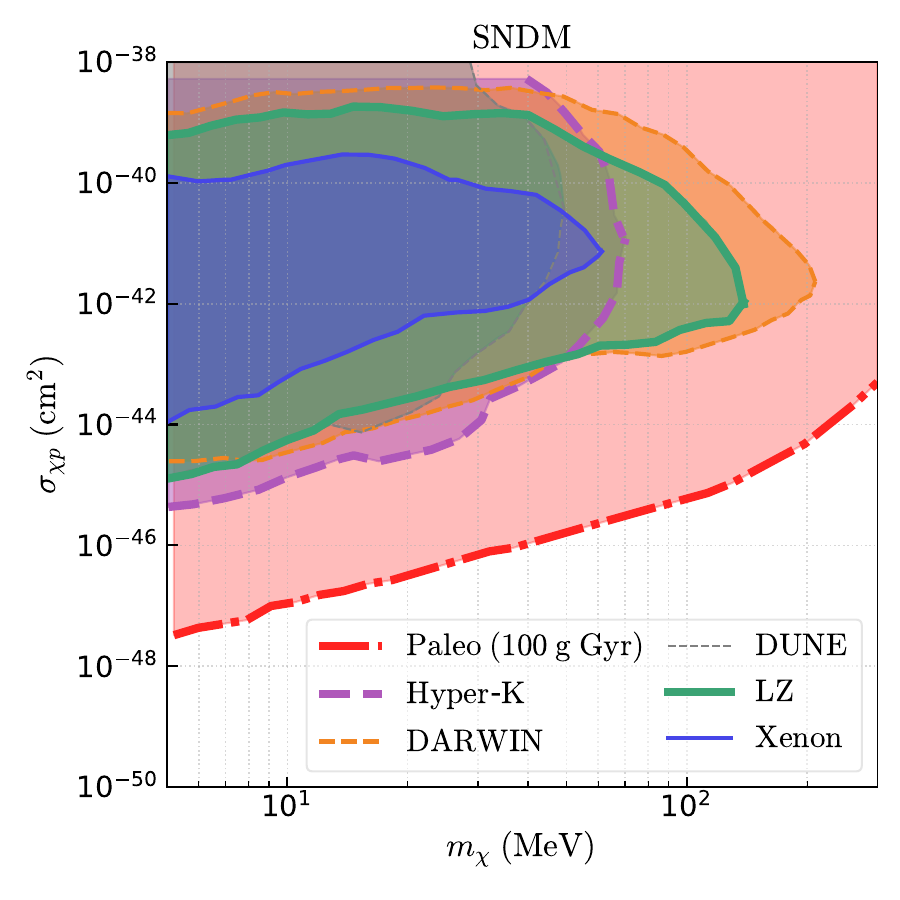}
 \caption{Projected sensitivities to the DM–nucleon scattering cross section for boosted DM.
\textbf{Left panel} constrains the spin-independent DM–nucleon cross section in the CRDM case. The dash-dotted red line shows the 95\% C.L.  sensitivity from paleo-detectors with an exposure of 100 g$\cdot$Gyr. 
The blue, cyan and purple solid lines show the limits from XENON1T~\cite{XENON:2018voc}, PandaX~\cite{PandaX-II:2021kai} and Super-K~\cite{Super-Kamiokande:2022ncz}, respectively. 
\textbf{Right panel} considers the SNDM mechanism using a dark photon model, and derive the corresponding sensitivity to  $\sigma_{\chi p}$ as defined by Eq.~\eqref{eq:effchip}. The dash-dotted red line again represents the paleo-detector sensitivity at 95\% C.L. with 100 g$\cdot$Gyr exposure. The purple, gray, and orange dashed lines show the sensitivities of Hyper-K~\cite{Hyper-Kamiokande:2021frf}, DUNE~\cite{DUNE:2020zfm}, and DARWIN~\cite{DARWIN:2016hyl}, respectively, and the green and blue solid lines show the limits of LZ~\cite{LZ:2018qzl} and XENON1T~\cite{XENON:2018voc}, respectively.
}\label{fig:limits}
 \end{figure*}

Our final results are presented as the red-shaded regions (with dash-dotted boundaries) in both panels of Fig.~\ref{fig:limits}, showing the 95\% C.L.  sensitivities of a paleo-detector with an exposure of $100$~g$\cdot$Gyr on the DM–nucleon scattering cross section for both the CRDM and SNDM mechanisms. The CRDM case (left panel) assumes a spin‑independent constant cross‑section. As already explained above, in this case the sensitivity of paleo-detector decreases with the DM mass, due to the fact that the local number density of heavier DM  is smaller. In contrast, the SNDM case (right panel) adopts a dark photon mediator framework, where the detectable parameter region is derived for the quantity $\sigma_{\chi p}$ defined by Eq.~\eqref{eq:effchip}. While the DM flux reaches its maximum at the transition between the free-streaming and trapping regimes, the lower boundary of the detectable parameter region usually stays below this transition, in the free-streaming regime. In this regime, the physics is usually insensitive to $m_\chi$ before the production of DM particles get Boltzmann-suppressed at $T\gg m_\chi$. Indeed, the apparent dependence on the DM mass  mostly comes from the $m_\chi^2$ factor in the definition of $\sigma_{\chi p}$. Consequently,  our projected sensitivities also apply for DM lighter than $5\,$MeV, where the  paleo-detectors are able to probe even smaller values of $\sigma^{SI}_{\chi p}$ ($\sigma_{\chi p}$) in the CRDM  (SNDM) case.

In addition, we include in the left panel some existing experiments, such as  the limits (solid lines)  from XENON1T~\cite{XENON:2018voc}, PandaX~\cite{PandaX-II:2021kai} and Super‑K~\cite{Super-Kamiokande:2022ncz}. 
Similarly, the right panel for the SNDM case also shows the limits (solid)  from XENON1T~\cite{XENON:2018voc} and PandaX~\cite{PandaX-II:2021kai}, as well as the projected sensitivities (dotted) of Hyper‑K~\cite{Hyper-Kamiokande:2021frf}, DUNE~\cite{DUNE:2020zfm}, DARWIN~\cite{DARWIN:2016hyl}.
As the plots explicitly show, paleo‑detectors will be able to provide the strongest constraints for DM masses of $\mathcal{O}(1)$ to $\mathcal{O}(100)$~MeV in both cases. 
Because particle models can be easily adjusted to make DM  lepto-phobic at leading order and we focus on the DM-nucleon interaction, direct detection experiments solely relying on DM-electron interactions are not considered in this work. 

Before  proceeding to the conclusion,  we briefly discuss the rationale for the axis value ranges employed in  Fig.~\ref{fig:limits}, while a detailed investigation of this aspect is beyond the scope of this work. In the CRDM case,  for $\sigma^{\rm}_{\chi p} \gtrsim 10^{-29}$\,cm$^2$ the DM flux would get quickly attenuated by scattering with   Earth before reaching the paleo-detector target. This has been illustrated in right panel of Fig.~\ref{fig:CRDM_flux}.  Similarly, for the SNDM mechanism  the trapping effect would become so significant for $\sigma_{\chi p} \gtrsim 10^{-38}$\,cm$^2$ that it can hardly be compensated by the enhanced scattering cross section between DM  and paleo-detectors. Clearly, for smaller scattering cross sections, of interest here, Earth attenuation of the DM flux is negligible and thus is not considered. Regarding the range choice of our $x$-axis, DM masses below 5\,MeV are already  very strongly constrained by early Universe observables; see \emph{e.g.} recently \cite{Sabti:2019mhn, Giovanetti:2021izc,Chu:2023jyb}. Thus we do not further get to even lighter DM,  where our results definitely apply.  At the large-mass end approaching GeV-scale, the SN production of the DM flux get exponentially suppressed as its mass exceeds the SN temperature.  Moreover,  direct detection experiments with low energy-threshold also start to play   important roles in this  mass range. such as  CRESST-III~\cite{CRESST:2019jnq,CRESST:2024cpr}, DarkSide-50~\cite{DarkSide-50:2022qzh}, PandaX-4T~\cite{PandaX:2023xgl}, and DAMIC-M~\cite{DAMIC-M:2025luv}.

\section{Conclusion}
\label{sec:Conclusion}
Astrophysical acceleration mechanisms have recently been invoked extensively to boost a fraction of DM particles with sub-GeV masses to higher kinetic energies.  Without such boosting, the nuclear recoil energy triggered by astrophysical DM particles lies below the typical experimental threshold in conventional direct detection experiments. The presence of certain acceleration mechanisms makes this mass range much easier to probe.  We thus  consider two such mechanisms to generate a large amount of (semi-)relativistic DM flux, CR-boosted DM with a constant spin‑independent cross section and SN-source DM with a dark photon mediator. Meanwhile, paleo-detectors can accumulate extremely high exposure over geological timescales, offering unprecedented sensitivity to rare signals. They are  well suited for detecting such boosted dark matter, especially when the boosting itself also occurred on geological timescales.

In this work we have carefully calculated the fluxes of both the CR-boosted and SN-sourced DM particles reaching paleo-detectors at Earth, and obtained the nuclear recoil track length distributions within olivine crystals. Following previous studies, we incorporate all relevant background components,  including solar neutrinos, SN neutrinos, atmospheric neutrinos, neutrons originating from uranium decay chains, and thorium recoils. With all uncertainties treated as Gaussian priors and associated nuisance parameters,  we have derived the 95\% C.L. detection sensitivity of paleo-detectors  to the DM-nucleon scattering cross section for sub-GeV DM. For both the CRDM and SNDM mechanisms,  paleo-detectors with an exposure of $100 \rm \;g \cdot Gyr$ provide stronger constraints than conventional DM and neutrino experiments in the mass range of $ \mathcal{O}(1) - \mathcal{O}(100) \mev$, owing to the  exceptionally high integrated exposure. 
Between the two, the SNDM scenario is particularly appealing. Not only can it lead to constraints on DM–nucleon interactions that exceed those from the CRDM one by more than ten orders of magnitude, but also, provided the DM mass lies below or around the supernova temperature, the production and emission of a DM flux from past supernova events are unavoidable.

In summary, paleo-detectors offer a powerful means to probe boosted sub-GeV DM,  as has been demonstrated in this work. Their ability to accumulate exposure over ancient timescales makes them uniquely capable of searching for rare signals from astrophysical DM sources in the past. At this moment, one should remain cautious about  the current understanding of paleo‑detector backgrounds. Nevertheless, with further improvements in track readout resolution and background reduction, paleo‑detectors are expected to become an important tool in the next‑generation search for sub-GeV DM interactions.

\section*{Acknowledgments}
M.W.Y and Y.L.S.T. are supported in part by 
the National Key Research and Development Program of China (grant No. 2022YFF0503304), the National Science Foundation of China (grant No. 12588101), the Project for Young Scientists in Basic Research of the Chinese Academy of Sciences (grant  No. YSBR-092), and 
the China Manned Space Program (grant No. CMS-CSST-2025-A03). X.C. is funded by the National Natural Science Foundation of China (grant No. E4146602 and No. 12547104), and the Fundamental Research Funds for the Central Universities (grant No. E4EQ6605X2 and E5ER6601A2). 


\appendix
\section{Robustness check with bin-wise background uncertainty}\label{Robustness check with bin-wise background uncertainty}

\begin{figure*}[t]
\centering
\includegraphics[width=8.85cm]{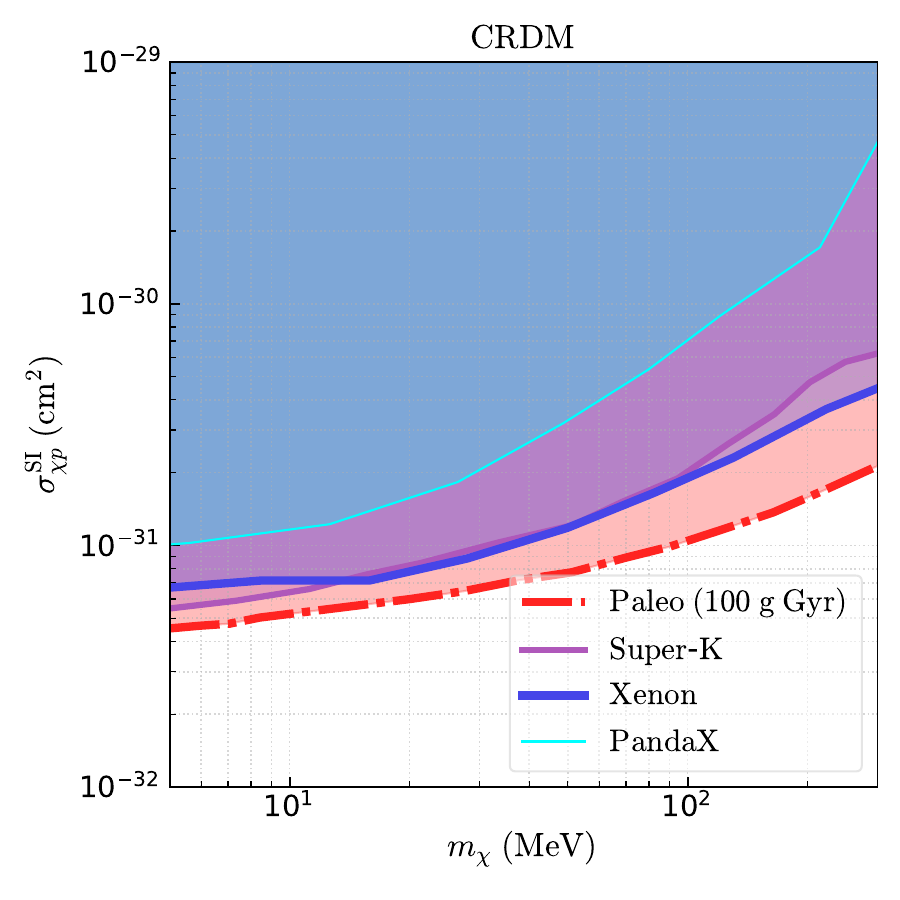}
\includegraphics[width=8.85cm]{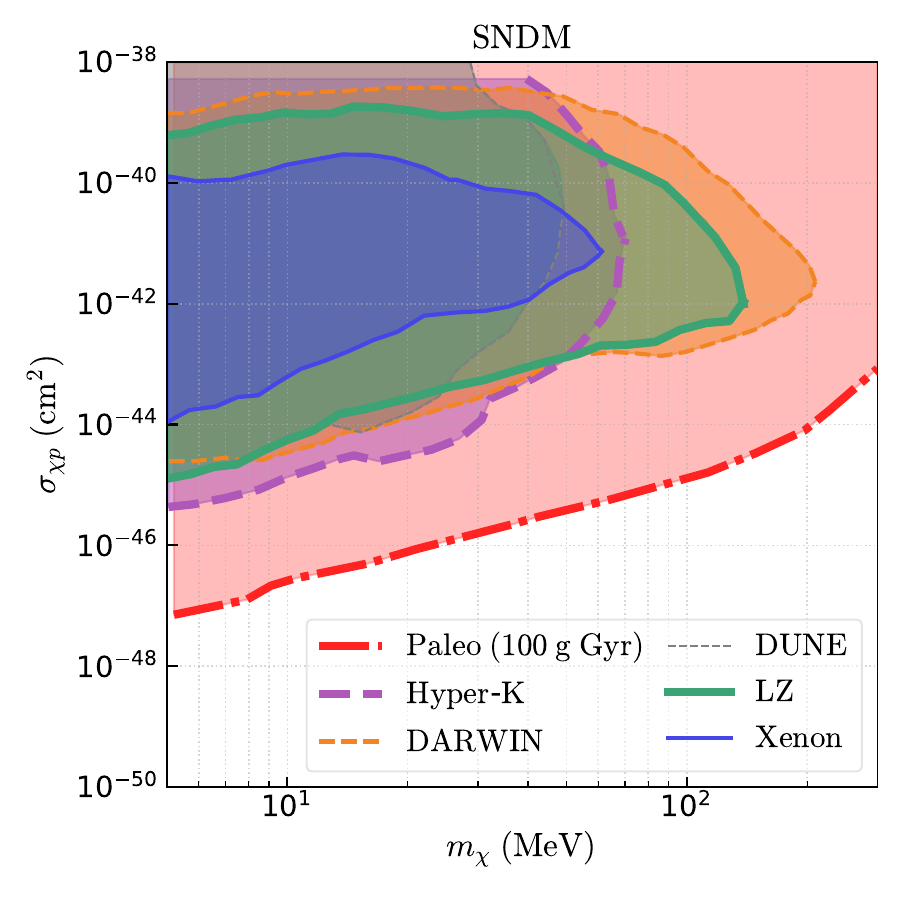}
 \caption{Robustness check of the projected sensitivities to the
DM--nucleon scattering cross section for boosted DM using the
conservative bin-wise variance prescription in Eq.~(\ref{chi2_cons}). The left panel
shows the CRDM case, while the right panel shows the SNDM case in the
dark-photon model. The
dash-dotted red lines denote the 95\% C.L. paleo-detector sensitivities
for an exposure of \(100~{\rm g\,Gyr}\). The existing limits and projected
sensitivities from other experiments are the same as those shown in
Fig.~\ref{fig:limits}.
}\label{fig:limits_cons}
 \end{figure*}

In the main analysis, the background uncertainties are implemented as
correlated normalization nuisance parameters for the individual
background components. This treatment preserves the spectral-shape
information in the binned track-length distribution. To assess the
impact of a more conservative uncertainty prescription, we also perform
a simplified robustness check in which an additional bin-wise variance is
assigned to the background prediction. We use
\begin{equation}\label{chi2_cons}
\chi^2_{\rm cons}
=
\sum_i
\frac{S_i^2}
{R_{{\rm data},i}+\left(0.01R_{{\rm data},i}\right)^2}\, ,
\end{equation}
where \(S_i\) is the signal count in the \(i\)-th bin and
\(R_{{\rm data},i}\) is the background-only Asimov count. This prescription
does not represent a full physical covariance model. It is instead used
as an illustrative test that weakens the spectral-shape information by
allowing an additional bin-wise uncertainty at the level of the
background rate.

The resulting projected sensitivities are shown in Fig.~\ref{fig:limits_cons}.
The limits become weaker by about a factor of two compared with the main
analysis. Nevertheless, the paleo-detector sensitivity remains stronger
than the current direct-detection limits in the parameter region shown.
A complete treatment of shape systematics would require a covariance
matrix or dedicated shape nuisance parameters for detector response. We leave this for future work.

\FloatBarrier
\bibliography{refs}
\end{document}